\documentclass[journal, 12pt, onecolumn, draftclsnofoot]{IEEEtran}
\usepackage{subcaption}
\usepackage{amsmath, amssymb}
\usepackage{xcolor}
\usepackage[pdftex]{graphicx}
\usepackage{tikz}
\usepackage[font=small,labelfont=bf]{caption}
\usetikzlibrary{positioning, arrows, calc}
\newcommand{\BitMask}{\emph{BitMask}}
\newcommand{\TopBits}{\emph{TopBits}}

\begin{document}
%
\title{Functional Error Correction for \\ Robust Neural Networks}
%
%
%

\author{Kunping Huang,
        Paul Siegel,
        Anxiao (Andrew) Jiang
\thanks{Kunping Huang and Anxiao (Andrew) Jiang are with the Department
of Computer Science and Engineering, Texas A\&M University, College Station,
TX, 77843 USA. E-mail: kun150kun@tamu.edu, ajiang@cse.tamu.edu.}
\thanks{Paul Siegel is with the Department of Electrical and Computer Engineer, University of California, San Diego, La Jolla, CA 92093 USA. E-mail: psiegel@ucsd.edu.}}

%
%

\markboth{Journal on Selected Areas in Information Theory}%
{Shell \MakeLowercase{\textit{et al.}}: Bare Demo of IEEEtran.cls for IEEE Journals}
%



\maketitle

\begin{abstract}

When neural networks (NeuralNets) are implemented in hardware, their weights need to be stored in memory devices. As noise accumulates in the stored weights, the NeuralNet's performance will degrade. This paper studies how to use error correcting codes (ECCs) to protect the weights. Different from classic error correction in data storage, the optimization objective is to optimize the NeuralNet's performance after error correction, instead of minimizing the Uncorrectable Bit Error Rate in the protected bits. That is, by seeing the NeuralNet as a function of its input, the error correction scheme is function-oriented. A main challenge is that a deep NeuralNet often has millions to hundreds of millions of weights, causing a large redundancy overhead for ECCs, and the relationship between the weights and its NeuralNet's performance can be highly complex. To address the challenge, we propose a Selective Protection (SP) scheme, which chooses only a subset of important bits for ECC protection. To find such bits and achieve an optimized tradeoff between ECC's redundancy and NeuralNet's performance, we present an algorithm based on deep reinforcement learning. Experimental results verify that compared to the natural baseline scheme, the proposed algorithm achieves substantially better performance for the functional error correction task.
\end{abstract}


%
\IEEEpeerreviewmaketitle

\section{Introduction}
\label{sec:intro}
Deep learning has become a boosting force for AI with many applications. When a neural network is implemented in hardware, its weights need to be stored in memory devices. Noise in such devices will accumulate over time, causing the neural network's performance to degrade. It is important to protect neural networks using error correction schemes. In this work, we study how to use error correcting codes (ECCs) to protect the weights of neural networks.

\begin{figure}[!t]
    \centering
    \begin{subfigure}[b]{0.45\textwidth}
        \includegraphics[width=\textwidth]{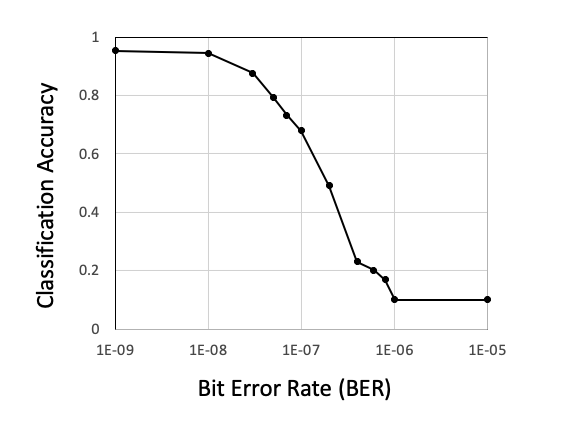}
        \subcaption{}
        \label{fig:noise_a}
    \end{subfigure}
    \hfill
    \begin{subfigure}[b]{0.45\textwidth}
        \includegraphics[width=\textwidth]{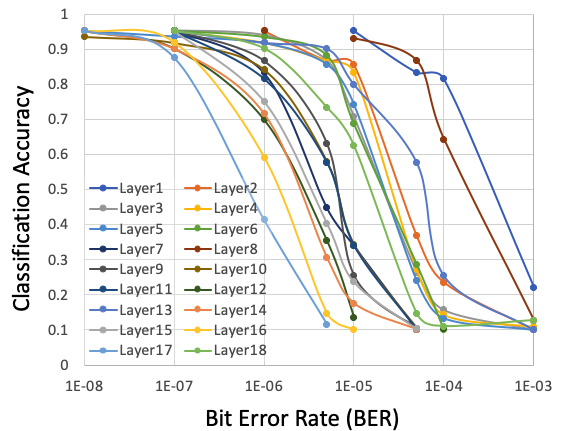}
        \subcaption{}
        \label{fig:layer_noise}
    \end{subfigure}
    \caption{The BER-performance tradeoff for a neural network. Here the network is ResNet-18 trained on the CIFAR-10 dataset. (a) The curve shows how the network's performance (classification accuracy) decreases as the bit error rate (BER) in the bits that represent the network's weights increases. (b) The curves show how errors in different layers of the neural network have different impact on the neural network's performance. Here we add errors to the weights of only one layer in the neural network.}
    \label{fig:noise}
\end{figure}

The protection of neural networks has a different optimization objective from classic error correction in data storage systems. In classic error correction, the objective is to minimize the Uncorrectable Bit Error Rate (UBER) in the protected bits. For neural networks, however, the objective is to optimize its performance (e.g., classification accuracy). That is, by seeing the neural network as a function of its input, the error correction scheme is function-oriented.

\begin{figure}[!t]
    \centering
    \begin{minipage}[b]{0.45\textwidth}
        \includegraphics[width=\textwidth]{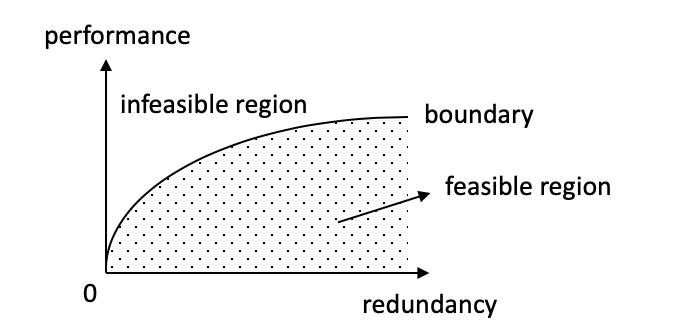}
        \caption{The redundancy-performance tradeoff for protecting a neural network. The boundary of the feasible region shows the optimal achievable performance of the neural network given the redundancy of ECC for protecting its weights.}
        \label{fig:rp}
    \end{minipage}
    \hfill
    \begin{minipage}[b]{0.45\textwidth}
        \includegraphics[width=0.9\textwidth]{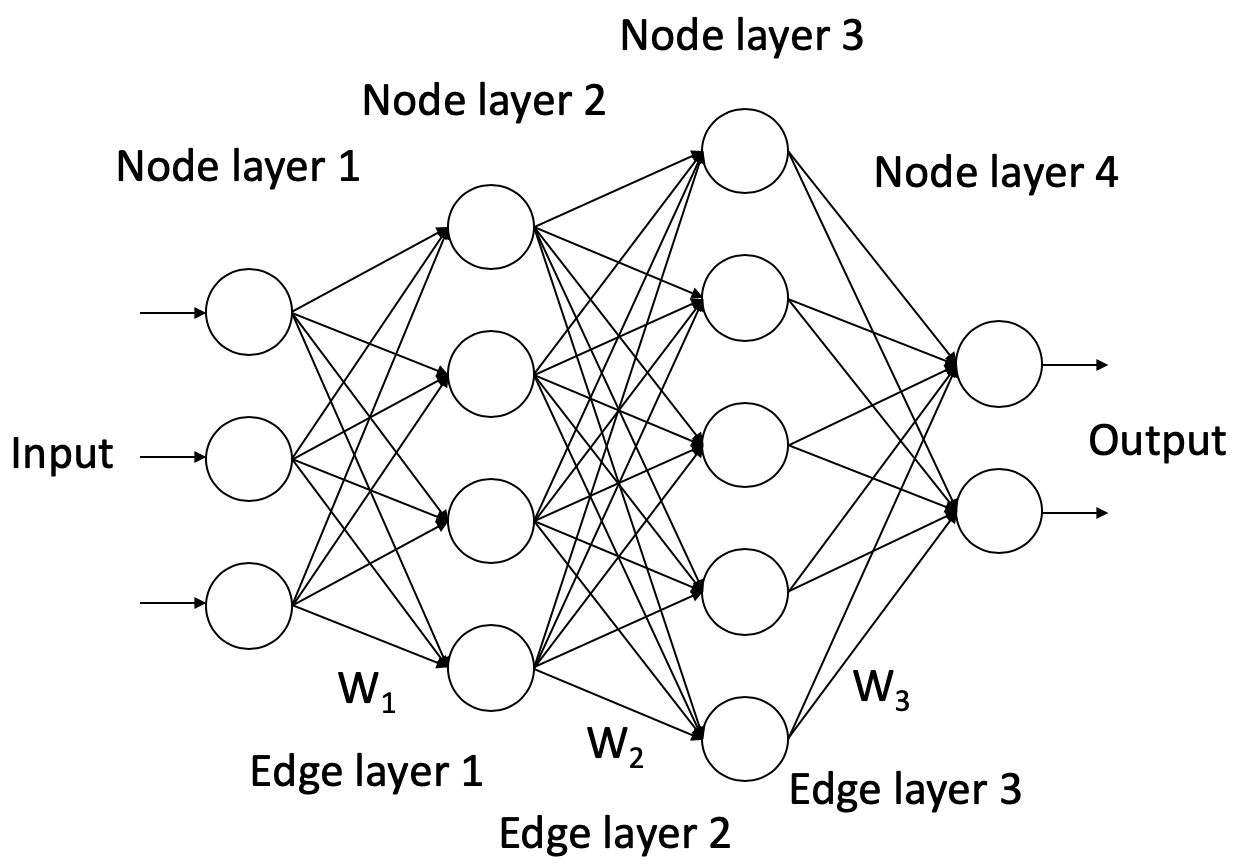}
        \caption{A neural network with four node layers (an input layer, two hidden layers and an output layer) and three edge layers. Here $W_{1}, W_{2}, W_{3}$ are the set of weights in each edge layer.}
        \label{fig:dnn}
    \end{minipage}
\end{figure}

Several challenges exist for the protection of neural networks. First of all, a deep neural network (DNN) often has many weights. For example, DNNs in computer vision often have millions to hundreds of millions of weights~\cite{DBLP:journals/corr/HeZRS15}. This can cause a very large redundancy overhead for ECCs. So it is important to design schemes that can reduce redundancy, and achieve an optimized redundancy-performance tradeoff. Such a tradeoff is illustrated in Figure~\ref{fig:rp}.

Secondly, the relationship between a neural network's weights and its performance is highly complex. Understanding on the relationship is very limited, and is an active topic of research in many areas~\cite{DBLP:journals/corr/KirkpatrickPRVD16, han2015deep}. Therefore, it is very challenging to design efficient algorithms that can identify weights that are most important for preserving the performance of neural networks.

We illustrate in Figure~\ref{fig:noise} how a neural network's performance is affected by noise in its weights. The network considered here is ResNet-18~\cite{DBLP:journals/corr/HeZRS15}, a well-known network for image classification. It consists of 19 layers of nodes and 26 layers of edges (including 8 layers of skip connections). Among the 26 edge layers, 21 of them have trainable weights. When binary-symmetric errors appear in the bits that represent the network's weights, the relation between the Bit Error Rate (BER) and the network's performance (i.e., classification accuracy) is shown in Figure~\ref{fig:noise} (a). (For a more detailed study on the relation between errors and neural networks' performance, see the nice work in~\cite{DBLP:journals/corr/abs-1709-06173}.) It can be seen that when the BER is quite small, the network's performance does not degrade much. However, once the BER exceeds a certain threshold, its performance starts to degrade substantially. This relation is common for various types of neural networks~\cite{DBLP:journals/corr/abs-1709-06173, eccnnn}. It implies that to protect a neural network, a good redundancy-performance tradeoff can be achieved by keeping the UBER below a certain threshold, especially for those bits that are most critical to the neural network's performance.

We further illustrate that the noise in different layers of a neural network has different impact on its performance. (Similar results have been shown in~\cite{DBLP:journals/corr/abs-1709-06173}.) We add noise to the weights of only one layer of edges in ResNet-18 at a time, and the result is shown in Figure~\ref{fig:noise} (b).~\footnote{For simplicity, the result here is only for the first 18 layers of edges with weights.} It can be seen that even for the same BER, different layers' noise can impact performance quite differently. Therefore, to optimize the redundancy-performance tradeoff, different layers should receive different levels of protection.

In this paper, we propose a Selective Protection (SP) scheme, which chooses only a subset of important bits for ECC protection. Furthermore, for different layers of edges, the numbers of protected bits for their weights are different. The scheme uses the fact that different layers impact performance differently. However, since layers jointly determine a network's performance in complex ways, when noise exists in all layers, how to optimize the scheme is still a challenging problem.

To address the challenge, we present an algorithm based on deep reinforcement learning. The key of the algorithm is to learn the complex relation between which bits to protect and the network's corresponding performance. That is, given the knowledge on which bits are protected from errors, we learn a function that can predict the performance of the neural network. We then use the prediction to optimize the set of protected bits, and then the network's corresponding true performance is measured as a feedback reward signal to help further refine the accuracy of the above performance-prediction function. The above learning process repeats itself until its performance converges. To reduce the complexity of learning, we decompose the above process by layers, where the network's layers sequentially take the actions of performance prediction and bit selection. Note that the bits selected for protection in each layer can be a mask vector instead of a single number, that is, we need to decide \emph{which} bits to protect instead of just \emph{how many} bits to protect. That is due to an interesting finding in this paper that, depending on how weights are represented as bits, those bits most worthy of protection are not necessarily the Most Significant Bits (MSBs). Furthermore, since we focus on optimizing the redundancy-performance tradeoff, the ECC redundancy is set as an integrated component in the reward function.

Our algorithm can be evaluated based on the redundancy-performance tradeoff as follows. Let $k_{total}$ denote the total number of bits used to represent the neural network's weights. Let $k_{pro}$ denote the number of bits we protect with ECCs. Let the ECCs be $(n, k)$ linear codes, where $n$ denotes the codeword length and $k$ denotes the number of information bits. Then the number of parity-check bits is $\frac{n-k}{k} \cdot k_{pro}$. We normalize it by $k_{total}$, and call it \emph{redundancy} $r$, namely,
\begin{equation}
    \label{eq:r}
    r = \frac{k_{pro} (n-k)}{k_{total} k}.
\end{equation}
As for the performance of the neural network, for classification tasks (which this work focuses on), it usually refers to the classification accuracy, namely, the probability that the inputs are classified correctly.

We compare the performance of our algorithm to a natural baseline scheme, where all layers of the neural network receive the same level of protection from ECCs. Experimental results verify that our proposed algorithm achieves substantially better performance. For example, when the neural network is ResNet-18 and its weights are represented by bits using the IEEE-754 standard (i.e., the single-precision floating-point format), and when BER is $1\%$, the baseline scheme's classification accuracy drops very quickly once its redundancy $r$ is below the threshold 0.04525. In comparison, our algorithm can decrease the corresponding threshold to 0.03879, which represents a reduction of 14.3\% in the redundancy requirement. If the ECC approaches the Shannon capacity, this reduction can be further enlarged to 25.7\%.

\begin{figure}
    \centering
    \begin{subfigure}[b]{0.45\textwidth}
    \centering
    \begin{tikzpicture}[scale=2, state/.style={rectangle, draw, minimum height=0.01cm,minimum width=3cm}, font=\scriptsize]
    \node[state, fill={rgb,255:red,250; green,235; blue,215}] (l1) {3$\times$3 conv, 64};
    \node[state, below = 0.2cm of l1, fill={rgb,255:red,250; green,235; blue,215}] (l2) {3$\times$3 conv, 64};
    \node[below = 0.2cm of l2] (l3) {pooling, /2};
    \node[state, below = 0.2cm of l3, fill={rgb,255:red,225; green,215; blue,235}] (l4) {3$\times$3 conv, 128};
    \node[state, below = 0.2cm of l4, fill={rgb,255:red,225; green,215; blue,235}] (l5) {3$\times$3 conv, 128};
    
    \node[below = 0.2cm of l5] (l6) {pooling, /2};
    \node[state, below = 0.2cm of l6, fill={rgb,255:red,200; green,240; blue,170}] (l7) {3$\times$3 conv, 256};
    \node[state, below = 0.2cm of l7, fill={rgb,255:red,200; green,240; blue,170}] (l8) {3$\times$3 conv, 256};
    \node[state, below = 0.2cm of l8, fill={rgb,255:red,200; green,240; blue,170}] (l9) {3$\times$3 conv, 256};
    
    \node[below = 0.2cm of l9] (l10) {pooling, /2};
    \node[state, below = 0.2cm of l10, fill={rgb,255:red,245; green,200; blue,185}] (l11) {3$\times$3 conv, 512};
    \node[state, below = 0.2cm of l11, fill={rgb,255:red,245; green,200; blue,185}] (l12) {3$\times$3 conv, 512};
    \node[state, below = 0.2cm of l12, fill={rgb,255:red,245; green,200; blue,185}] (l13) {3$\times$3 conv, 512};
    
    \node[below = 0.2cm of l13] (l14) {pooling, /2};
    \node[state, below = 0.2cm of l14, fill={rgb,255:red,190; green,225; blue,240}] (l15) {3$\times$3 conv, 512};
    \node[state, below = 0.2cm of l15, fill={rgb,255:red,190; green,225; blue,240}] (l16) {3$\times$3 conv, 512};
    \node[state, below = 0.2cm of l16, fill={rgb,255:red,190; green,225; blue,240}] (l17) {3$\times$3 conv, 512};
    
    \node[below = 0.2cm of l17] (l18) {pooling};
    \node[state, below = 0.2cm of l18] (l19) {fc-512};
    \node[state, below = 0.2cm of l19] (l20) {fc-512};
    \node[state, below = 0.2cm of l20] (l21) {fc-512};
    
    \draw [->,>=stealth] (l1) -- (l2);
    \draw [->,>=stealth] (l2) -- (l3);
    \draw [->,>=stealth] (l3) -- (l4);
    \draw [->,>=stealth] (l4) -- (l5);
    \draw [->,>=stealth] (l5) -- (l6);
    \draw [->,>=stealth] (l6) -- (l7);
    \draw [->,>=stealth] (l7) -- (l8);
    \draw [->,>=stealth] (l8) -- (l9);
    \draw [->,>=stealth] (l9) -- (l10);
    \draw [->,>=stealth] (l10) -- (l11);
    \draw [->,>=stealth] (l11) -- (l12);
    \draw [->,>=stealth] (l12) -- (l13);
    \draw [->,>=stealth] (l13) -- (l14);
    \draw [->,>=stealth] (l14) -- (l15);
    \draw [->,>=stealth] (l15) -- (l16);
    \draw [->,>=stealth] (l16) -- (l17);
    \draw [->,>=stealth] (l17) -- (l18);
    \draw [->,>=stealth] (l18) -- (l19);
    \draw [->,>=stealth] (l19) -- (l20);
    \draw [->,>=stealth] (l20) -- (l21);
    \end{tikzpicture}
    \end{subfigure}
    \begin{subfigure}[b]{0.45\textwidth}
    \centering
    \begin{tikzpicture}[scale=2, state/.style={rectangle, draw, minimum height=0.01cm,minimum width=3cm}, font=\scriptsize]
    \node[state, fill={rgb,255:red,250; green,235; blue,215}] (l1) {3$\times$3 conv, 64};
    \node[state, below = 0.4cm of l1, fill={rgb,255:red,225; green,215; blue,235}] (l2) {3$\times$3 conv, 64};
    \node[state, below = 0.2cm of l2, fill={rgb,255:red,225; green,215; blue,235}] (l3) {3$\times$3 conv, 64};
    \node[state, below = 0.2cm of l3, fill={rgb,255:red,225; green,215; blue,235}] (l4) {3$\times$3 conv, 64};
    \node[state, below = 0.2cm of l4, fill={rgb,255:red,225; green,215; blue,235}] (l5) {3$\times$3 conv, 64};
    
    \node[state, below = 0.2cm of l5, fill={rgb,255:red,200; green,240; blue,170}] (l6) {3$\times$3 conv, 128, /2};
    \node[state, below = 0.2cm of l6, fill={rgb,255:red,200; green,240; blue,170}] (l7) {3$\times$3 conv, 128};
    \node[state, below = 0.2cm of l7, fill={rgb,255:red,200; green,240; blue,170}] (l8) {3$\times$3 conv, 128};
    \node[state, below = 0.2cm of l8, fill={rgb,255:red,200; green,240; blue,170}] (l9) {3$\times$3 conv, 128};
    
    \node[state, below = 0.2cm of l9, fill={rgb,255:red,245; green,200; blue,185}] (l10) {3$\times$3 conv, 256, /2};
    \node[state, below = 0.2cm of l10, fill={rgb,255:red,245; green,200; blue,185}] (l11) {3$\times$3 conv, 256};
    \node[state, below = 0.2cm of l11, fill={rgb,255:red,245; green,200; blue,185}] (l12) {3$\times$3 conv, 256};
    \node[state, below = 0.2cm of l12, fill={rgb,255:red,245; green,200; blue,185}] (l13) {3$\times$3 conv, 256};
    
    \node[state, below = 0.2cm of l13, fill={rgb,255:red,190; green,225; blue,240}] (l14) {3$\times$3 conv, 512, /2};
    \node[state, below = 0.2cm of l14, fill={rgb,255:red,190; green,225; blue,240}] (l15) {3$\times$3 conv, 512};
    \node[state, below = 0.2cm of l15, fill={rgb,255:red,190; green,225; blue,240}] (l16) {3$\times$3 conv, 512};
    \node[state, below = 0.2cm of l16, fill={rgb,255:red,190; green,225; blue,240}] (l17) {3$\times$3 conv, 512};
    
    \node[below = 0.2cm of l17] (l18) {average pooling};
    \node[state, below = 0.2cm of l18] (l19) {fc-1000};
    
    \draw [->,>=stealth] (l1) -- (l2);
    \draw [->,>=stealth] (l2) -- (l3);
    \draw [->,>=stealth] (l3) -- (l4);
    \draw [->,>=stealth] (l4) -- (l5);
    \draw [->,>=stealth] (l5) -- (l6);
    \draw [->,>=stealth] (l6) -- (l7);
    \draw [->,>=stealth] (l7) -- (l8);
    \draw [->,>=stealth] (l8) -- (l9);
    \draw [->,>=stealth] (l9) -- (l10);
    \draw [->,>=stealth] (l10) -- (l11);
    \draw [->,>=stealth] (l11) -- (l12);
    \draw [->,>=stealth] (l12) -- (l13);
    \draw [->,>=stealth] (l13) -- (l14);
    \draw [->,>=stealth] (l14) -- (l15);
    \draw [->,>=stealth] (l15) -- (l16);
    \draw [->,>=stealth] (l16) -- (l17);
    \draw [->,>=stealth] (l17) -- (l18);
    \draw [->,>=stealth] (l18) -- (l19);
    \draw [->, bend left=90, distance=2cm, thick] ($(l1)!0.5!(l2)$) to  ($(l3)!0.5!(l4)$);
    \draw [->, bend left=90, distance=2cm, thick] ($(l3)!0.5!(l4)$) to  ($(l5)!0.5!(l6)$);
    \draw [->, bend left=90, distance=2cm, thick, dash pattern={on 5pt off 2pt }] ($(l5)!0.5!(l6)$) to  ($(l7)!0.5!(l8)$);
    \draw [->, bend left=90, distance=2cm, thick] ($(l7)!0.5!(l8)$) to  ($(l9)!0.5!(l10)$);
    \draw [->, bend left=90, distance=2cm, thick, dash pattern={on 5pt off 2pt }] ($(l9)!0.5!(l10)$) to  ($(l11)!0.5!(l12)$);
    \draw [->, bend left=90, distance=2cm, thick] ($(l11)!0.5!(l12)$) to  ($(l13)!0.5!(l14)$);
    \draw [->, bend left=90, distance=2cm, thick, dash pattern={on 5pt off 2pt }] ($(l13)!0.5!(l14)$) to  ($(l15)!0.5!(l16)$);
    \draw [->, bend left=90, distance=2cm, thick] ($(l15)!0.5!(l16)$) to  ($(l17)!0.5!(l18)$);
    \end{tikzpicture}
    \end{subfigure}
    \caption{The architecture of VGG16 and ResNet-18 models. The left model is the VGG16 neural network~\cite{simonyan2014very}, and the right model is the ResNet-18 neural network~\cite{DBLP:journals/corr/HeZRS15}.}
    \label{fig:network}
\end{figure}
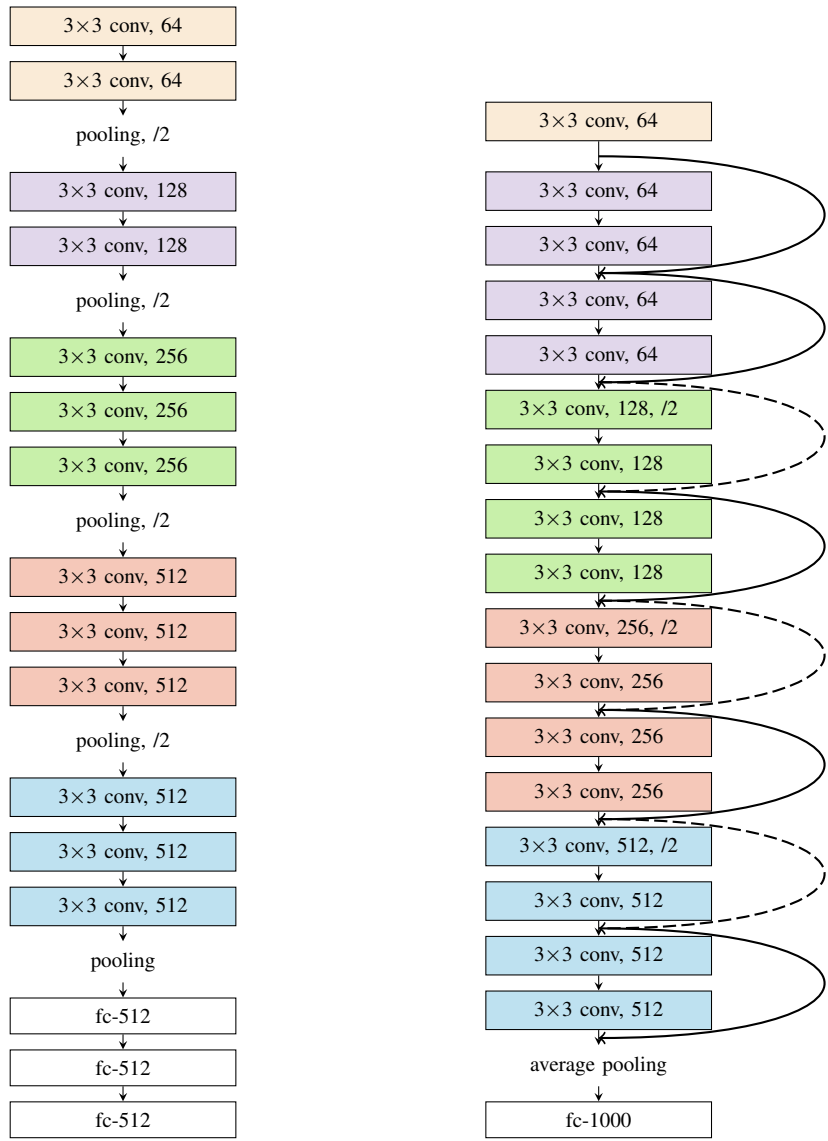

The rest of the paper is organized as follows. In Section~\ref{sec:rw}, we review related works. In Section~\ref{sec:drl}, we introduce the SP scheme, and present its deep reinforcement learning algorithm. In Section~\ref{sec:exp}, we evaluate the SP scheme by experiments, which verify that the scheme can substantially improve the redundancy-performance tradeoff for neural networks. The results also show that interestingly, depending on how weights are represented as bits, the bits that are most important to protect are not necessarily MSBs in the data representation. We present a detailed analysis for this interesting phenomenon. In Section~\ref{sec:con}, we present concluding remarks.

\section{Overview of Related Works}
\label{sec:rw}
The topic explored in this paper is related to several research areas. They include robustness of neural networks against noise, model compression, and reliability of computational circuits.

In the area of \emph{robustness of neural networks against noise}, researchers have studied the effect of noise on the performance of neural networks. In~\cite{DBLP:journals/corr/abs-1709-06173}, Qin \textit{et al.} studied random bit errors for weights stored as bits, and developed an ECC with one parity bit to improve the network's performance and robustness. In~\cite{echidnn}, Upadhyaya \textit{et al.} studied random noise for weights stored as analog numbers, and developed analog ECCs to correct the analog noise. In~\cite{8203770, 2019arXiv190312269S}, several security attack methods were tested to find specific error patterns that can cause serious damage to neural networks' performance. Note that different from the above works, this paper proposes the Selective Protection scheme for the first time, which protects different sets of bits for different layers. The scheme needs to protect all bits that are critical to the neural network's performance, not just bits that constitute a specific damaging error pattern.

In the area of \emph{model compression}, plenty of works have focused on how to reduce the size of a neural network without affecting its performance~\cite{han2015deep, luo2017thinet, DBLP:journals/corr/abs-1802-03494, DBLP:journals/corr/abs-1811-08886}. They use various techniques to either prune or quantize the weights in neural networks, and the simplified networks need to be retrained. Deep reinforcement learning methods, including the layer-by-layer training method, have been presented~\cite{DBLP:journals/corr/abs-1802-03494, DBLP:journals/corr/abs-1811-08886}. Note that in our work, we find important bits and protect them, without the need to modify the weights or retrain the network.

In the area of \emph{reliability of computational circuits}, researchers have studied the use of ECCs to ensure the correctness of circuits~\cite{514728, 4403210, 335020}. In comparison, our work focuses on the redundancy-performance tradeoff, where the neural network's performance does not have to be the same before and after ECC protection.

\section{Selective Protection Scheme by Deep Reinforcement Learning}
\label{sec:drl} 
In this section, we present the Selective Protection (SP) scheme for functional error correction. It protects the most important bits in weights by ECC in order to achieve an optimized redundancy-performance tradeoff. We first introduce weight representation for neural networks, and define the Selective Protection scheme. We then present a deep reinforcement learning (DRL) algorithm for the SP scheme.

\subsection{Weight Representation in Neural Networks}
\label{subsec:weight}
Neural networks have been used widely in deep learning. An example of a neural network is shown in Figure~\ref{fig:dnn}, which has four node layers and three edge layers between them. Examples of more complex neural networks, including VGG16 and ResNet-18, are shown in Figure~\ref{fig:network}. (Those two networks are important models for computer vision, and will be used in our experiments.) For ResNet-18, the skip connections between two node layers are also considered an edge layer.

There are different ways to represent weights in neural networks as bits. We introduce two important weight representations below. Both of them will be used in experiments.

\subsubsection{Standard Floating-Point Representation}
IEEE-754 is an international standard for floating-point representation. We adopt its 32-bit version. Given a weight $w \in \mathbb{R}$, let $B_{w}^{32} = (b_{0}, b_{1}, \cdots, b_{31})$ be its binary representation:
\begin{equation}
    w = (-1)^{(b_{0})_{2}} \times 2^{(b_{1}b_{2}\cdots b_{8})_{2}-127} \times (1.b_{9}b_{10} \cdots b_{31})_{2}
\end{equation}
Here $b_{0}$ is the \emph{sign bit}, $b_{1}b_{2}\cdots b_{8}$ are the \emph{exponent bits}, and $b_{9}b_{10} \cdots b_{31}$ are the \emph{fraction bits}. For example, if $B_{w}^{32} = (00111100001100000000000000000000)$, then $w = (-1)^{(0)_{2}} \times 2^{(01111000)_{2}-127} \times (1.01100000000000000000000)_{2} = (-1)^{0} \times 2^{120-127} \times 1.375 = 0.0107421875$. The IEEE-754 standard can represent values between $-2^{127}$ and $2^{127}$.

\subsubsection{Fixed-Point Representation}
In this representation, the weights in a range $[-c, c]$ are linearly quantized and represented as bits. (Such a representation has been used in neural networks before, including~\cite{DBLP:journals/corr/abs-1811-08886}.) Consider its $m$-bit version. Let $s = c / (2^{m - 1} - 1)$ be a scaling factor. Given a weight $w \in [-c, c]$, let $D_{w}^{m} = (b_{0}, b_{1}, \cdots, b_{m-1})$ be its binary representation:
\begin{equation}
    w = (-1)^{(b_{0})_{2}} \times (b_{1}b_{2}\cdots b_{m-1})_{2} \times s
\end{equation}
For example, when $c = 127$ and $m = 8$, if $D_{w}^{m} = (10010011)$, then $w = (-1)^{(1)_{2}} \times (0010011)_{2} \times (127 / (2^{8-1} - 1)) = (-1)^{1} \times 19 \times 1 = -19$.

\subsection{Selective Protection Scheme}
We now present the Selective Protection (SP) scheme, which selects important bits and protects them from errors with ECCs. Consider a neural network with $N$ edge layers. (In this paper, we consider error protection for weights on edges, not biases in nodes, because biases can often be implemented in alternative ways in hardware. Note that edge weights constitute by far the majority of all weights, and the results here can be naturally extended to biases as well.) For $i = 1, 2, \cdots, N$, let $L_{i}$ denote the $i$th edge layer, and let $W_{i}$ denote the set of weights in $L_{i}$. Assume that every weight is represented by $m$ bits. The SP scheme will select a \emph{bit-mask vector}
\begin{equation}
    M_{i} = (\mu_{i,0}, \mu_{i,1}, \cdots, \mu_{i, m-1}) \in \{0, 1\}^{m}
\end{equation}
for each edge layer $L_{i}$. For each weight $w = (b_{0}, b_{1}, \cdots, b_{m-1}) \in W_{i}$, its $j$th bit $b_{j}$ will be protected by ECC if $\mu_{i,j} = 1$. Naturally, we let $\mu_{i, j} = 1$ for the layer $L_{i}$ if its bits in the $j$th position are critical for the neural network's performance.

Note that the SP scheme applies the same bit-mask vector for all the weights in the same layer. In principle, every weight can be assigned its own bit-mask vector, but that will greatly increase the overhead of the scheme. By using one bit-mask vector per layer, a good balance between performance and overhead can be achieved.

The neural network has $k_{total} = m \sum_{i=1}^{N} |W_{i}|$ bits in total. The number of bits protected by ECCs is $k_{pro} = \sum_{i=1}^{N} |W_{i}| \sum_{j=0}^{m-1} \mu_{i,j}$. When the ECCs are $(n, k)$ linear codes, by Equation (\ref{eq:r}), the \emph{redundancy} of the SP scheme is
\begin{equation}
    r(M_{1}, M_{2}, \cdots, M_{N}) = \frac{(n-k) \sum_{i=1}^{N} |W_{i}| \sum_{j=0}^{m-1} \mu_{i,j}}{k m \sum_{i=1}^{N} |W_{i}|}
\end{equation}
Let $\mathcal{P}(M_{1}, M_{2}, \cdots , M_{N})$ denote the performance of the neural network (e.g. classification accuracy). Let $\bar{r}$ be a target redundancy. The optimization objective of SP scheme is to maximize $\mathcal{P}(M_{1}, M_{2}, \cdots , M_{N})$ given that $r(M_{1}, M_{2}, \cdots, M_{N}) = \bar{r}$. That is, after the ECCs are chosen appropriately based on the target Bit Error Rate, the SP scheme can be formulated as
\begin{equation}
    \begin{array}{ll}
        \text{max} & \mathcal{P}(M_{1}, M_{2}, \cdots , M_{N})\\
        \text{s.t.} & r(M_{1}, M_{2}, \cdots, M_{N}) = \bar{r}
    \end{array}
\end{equation}

\subsection{Deep Reinforcement Learning for Selective Protection}
We now present a deep reinforcement learning algorithm for the SP scheme. We assume that the bits suffer from errors of a Binary Symmetric Channel (BSC) with Bit Error Rate (BER) $p$, and a suitable $(n, k)$ linear ECC is used that can correct error of BER $p$ with a probability that approaches 1. Therefore, after error correction, only the bits not protected by ECC will have errors. Note that for a neural network, its performance is a highly complex function of its weights. The DRL algorithm will learn this complex function, and choose the important bits to protect accordingly.

In the following, we first present the essential components of the DRL algorithm: its \emph{state space}, \emph{action space}, \emph{reward function}, and \emph{policy of agents}. We then present the overall learning process of the DRL algorithm. 

\subsubsection{State Space}
There are two types of state spaces in our DRL algorithm: a \emph{Global State Space} and a set of \emph{Local State Spaces}. The global state space uses a set of parameters $\Theta$ to characterize the global configuration of the neural network. For $i = 1, 2, \cdots, N$, the $i$th edge layer has a local state space $\Pi_{i} \subset \Theta$, which is a partial view of the global state space used by the agent of the $i$th edge layer to take actions. Note that the parameters in $\Theta$ depend on the types of layers in the neural network. In our study, we focus on VGG16 and ResNet, which have two types of layers: convolutional layers and fully-connected layers. Therefore, the parameters in $\Theta$ are set accordingly, although they can be adjusted if other types of layers are considered. Note that a fully-connected layer can be seen as a special case of a convolutional layer, where its convolutional kernel has the same size as its input feature map.

For $i = 1, 2, \cdots, N$, let $c_{in}^{i}$ be the number of input channels for the $i$th layer $L_{i}$ (i.e., the number of input feature maps). Let $c_{out}^{i}$ be its number of output channels (i.e., the number of output feature maps). Let $s_{kernel}^{i}$ be its kernel size (i.e. the size of its filter for the convolution operation). Let $s_{stride}^{i}$ be its stride for convolution. Let $s_{feat}^{i}$ be the size of its input feature map (i.e., each input feature map is a  two-dimentional array of size $s_{feat}^{i} \times s_{feat}^{i}$). Let $a_{i} \in \mathcal{A}$ be the most recent action taken by the agent for $L_{i}$, where $\mathcal{A}$ denotes the action space, whose details will be introduced later. Let $\alpha_{i} = (c_{in}^{i}, c_{out}^{i}, s_{kernel}^{i}, s_{stride}^{i}, s_{feat}^{i}, |W_{i}|, a_{i})$ denote a state vector associated with $L_{i}$. Then, the global state $\theta \in \Theta$ is defined as
\begin{equation}
    \theta = (\alpha_{1}, \alpha_{2}, \cdots, \alpha_{N})
\end{equation}

To simplify the learning process, each layer $L_{i}$ uses a local state $\pi_{i} \in \Pi_{i}$ defined as follows:
\begin{equation}
\pi_{i} = (c_{in}^{i}, c_{out}^{i}, s_{kernel}^{i}, s_{stride}^{i}, s_{feat}^{i}, |W_{i}|, a_{i-1})
\end{equation}
When $i=1$, the parameter $a_{i-1} = a_{0}$ can be a constant. Note that in $\pi_{i}$, only the action of its previous layer $a_{i-1}$ is used, instead of the actions of all its previous layers $a_{1}, a_{2}, \cdots, a_{i-1}$.

\subsubsection{Action Space}
\label{subsubsect:as}
We now present the space of actions for the DRL algorithm. For $i = 1, 2, \cdots, N$, the action of the $i$th layer $L_{i}$ is to choose a value $a_{i} \in \{0, 1\}^{m}$ for its bit-mask vector $M_{i} = (\mu_{i,0}, \mu_{i,1}, \cdots, \mu_{i, m-1})$. The overall action is the sequence of actions $(a_{1}, a_{2}, \cdots, a_{N})$. Note that in each iteration of the DRL algorithm, the actions $a_{1}, a_{2}, \cdots, a_{N}$ are chosen sequentially. When the layer $L_{i}$ takes the action $a_{i}$, it chooses the value of $a_{i}$ (i.e., sets its bit-mask vector $M_{i}$) based on its local state $\pi_{i}$ and the reward function (to be introduced later).

Let the above method be called the \BitMask{} method. To make the method satisfy the redundancy constraint, the reward function not only considers the performance of the neural network, but also the distance between the current redundancy $r$ and the target redundancy $\bar{r}$. The reward value is actually a linear combination of the two terms. When the DRL algorithm ends, the final redundancy $r$ will be close, but not necessarily equal, to $\bar{r}$. By making the coefficient for the distance between $r$ and $\bar{r}$ sufficiently large in the reward value, we can make $r$ sufficiently close to $\bar{r}$.

We now present a simplified version of the \BitMask{} method, which we called the \TopBits{} method. In the \TopBits{} method, each layer always chooses the first few bits of its weights for ECC protection. (The number of bits chosen by different layers can still be different.) This method is intuitively understandable for the fixed-point representation, because the first bit $b_{0}$ is the sign bit (thus very important), and for the remaining bits, the More Significant Bits (MSBs) affect the value of the weight more significantly than the Less Significant Bits (LSBs). Similarly, for the IEEE-754 floating-point representation, the first bit $b_{0}$ is also the sign bit (thus important), the exponent bits (which follow $b_{0}$) affect the weight more significantly than the fraction bits, and the MSBs in the fraction bits affect the weight more significantly than LSBs. Therefore, it seems natural for the SP scheme to always protect the first few bits. The \TopBits{} method also simplifies the learning process compared to the \BitMask{} method. However, our study will show the surprising result that the \BitMask{} method can sometimes outperform the \TopBits{} method (namely, MSBs do not always affect the performance of neural networks more substantially than LSBs).

In the \TopBits{} method, the reward function considers only the performance of the neural network, and does not consider the distance between the current redundancy $r$ and the target redundancy $\bar{r}$. To satisfy the redundancy constraint, the method takes two rounds of actions across all the layers in each iteration of the DRL algorithm: 
\begin{itemize}
    \item In the first round, the $N$ layers take actions $(a_{1}, a_{2}, \cdots a_{N})$ sequentially. For $i = 1, 2, \cdots, N$, the action of the $i$th layer $L_{i}$ is to choose a value $a_{i} \in \{0, 1, \cdots, m \}$, and set the first $a_{i}$ bits of the bit-mask vector $M_{i}$ to $1$ and set its other bits to $0$. Namely, $L_{i}$ selects the first $a_{i}$ bits of each weight for ECC protection.
    \item In the second round, if the current redundancy $r$ is greater than the target redundancy $\bar{r}$, then for $i = 1, 2, \cdots, N$, each layer $L_{i}$ decreases its $a_{i}$ by 1 (but without making $a_{i}$ negative) and adjusts its $M_{i}$ accordingly. The layers take the above actions sequentially, and stop as soon as we have $r \leq \bar{r}$.
\end{itemize}

\subsubsection{Reward Function}
\label{subsubsec:rs}
We now present the reward function for the DRL algorithm. Let $\mathcal{P}_{0}$ describe the performance (e.g., classification accuracy) of the neural network without any bit errors. After each iteration of the DRL algorithm (where the $N$ layers take their actions $(a_{1}, a_{2}, \cdots, a_{N})$ and set their bit-mask vectors $(M_{1}, M_{2}, \cdots, M_{N})$ accordingly), random bit errors of BER $p$ are added to all bits in the $N$ layers (but note that some of them are chosen to be protected by ECCs), and then the performance $\mathcal{P}$ of the neural network is measured. For the \TopBits{} method, the reward function after the iteration is set as
\begin{equation}
    R_{\textit{TopBits}} = \mathcal{P} - \mathcal{P}_{0}
\end{equation}

For the \BitMask{} method, its reward function also needs to consider the distance between the redundancy $r$ after the iteration and the target redundancy $\bar{r}$. Let $\beta^{+}$ and $\beta^{-}$ to be two positive real numbers. We define a function $f(r, \bar{r})$ as:
\begin{equation}
    f(r, \bar{r}) = 
    \left \{ \begin{array}{l l}
            \beta^{+} (\bar{r} - r) & \text{if  } r \geq \bar{r} \\
            \beta^{-} (r - \bar{r}) & \text{if  } r < \bar{r}
          \end{array}
    \right .
\end{equation}
and define the reward function as:
\begin{equation}
    R_{\textit{BitMask}} = \mathcal{P} - \mathcal{P}_{0} + f(r, \bar{r})
\end{equation}
Note that $f(r, \bar{r}) \leq 0$, which represents a penalty for the reward function when the current redundancy $r$ deviates from the target redundancy $\bar{r}$. When $r \geq \bar{r}$ (an undesirable case because the current redundancy is too large), the penalty $\beta^{+} (\bar{r} - r)$ helps the DRL algorithm reduce the redundancy in the next iteration. When $r < \bar{r}$ (a desirable case because the current redundancy is sufficiently small), interestingly, it is also helpful to set a small penalty $\beta^{-} (r - \bar{r})$, because it can prevent the neural network from getting stuck in states of very low redundancy in the practical implementation of the DRL algorithm. We usually make $\beta^{-}$ much less than $\beta^{+}$. For example, we can set $\beta^{+} = 1$ and $\beta^{-} = 0.05$.

\subsubsection{Policy of Agents and the Learning Process}
In the DRL algorithm, every layer $L_{i}$ has an \emph{agent} $A_{i}$ that takes the action $a_{i}$ based on the local state $\pi_{i}$ and an estimated reward function $\hat{R}$. How the agent $A_{i}$ chooses the action $a_{i}$ based on the available information is called its \emph{policy}. In this part, we present the policy of the $N$ agents $A_{1}, A_{2}, \cdots, A_{N}$.

\begin{figure}[!t]
    \centering
    \includegraphics[width=0.7\textwidth]{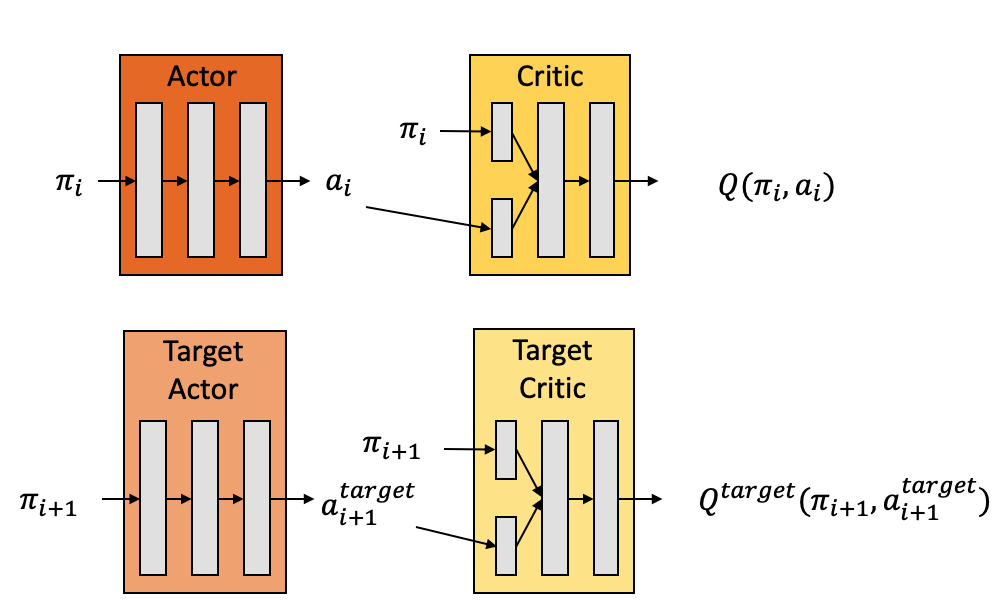}
    \caption{The four neural networks used in the deep reinforcement learning algorithm: the Actor Network (top left), the Target Actor Network (bottom left), the Critic Network (top right) and the Target Critic Network (bottom right).}
    \label{fig:actor_critic}
\end{figure}

We build four deep neural networks: an \emph{Actor Network}, a \emph{Target Actor Network}, a \emph{Critic Network}, and a \emph{Target Critic Network}. The four networks are illustrated in Figure~\ref{fig:actor_critic}. They are all Multilayer Perceptron (MLP) neural networks of four node layers, where the two hidden layers have size 400 and 300, respectively. Additional information on their architectures is as follows:
\begin{itemize}
    \item \emph{Actor Network} and \emph{Target Actor Network}: For both networks, the input is the local state $\pi_{i}$, and the output is the action $a_{i}$. The two networks have similar functions, but update their weights with different algorithms during training.
    \item \emph{Critic Network} and \emph{Target Critic Network}: For both networks, the input consists of the local state $\pi_{i}$ and the action $a_{i}$, and the output is an estimated value for the summation of the current and the future rewards in the same iteration (where future rewards are discounted in certain ways). Specifically, let $\gamma$ be a discount factor. Then for $t = 1, 2, \cdots, N $, the output of the two networks is the value of the following $Q$ function:
    \begin{equation}
        Q(\pi_{t}, a_{t}) = \sum_{i=t}^{N}\gamma^{i-t} \hat{R}(\pi_{i}, a_{i})
    \end{equation}
    where $\hat{R}(\pi_{i}, a_{i})$ is an estimation of the real reward of this iteration. As before, the two networks also have similar functions, but update their weights differently during training.
\end{itemize}
The DRL algorithm keeps using the Actor Network to generate actions. In each iteration, the $N$ agents $A_{1}, A_{2}, \cdots, A_{N}$ generate the actions $a_{1}, a_{2}, \cdots, a_{N}$ sequentially. That is, for $i=1, 2, \cdots, N$, the Actor Network takes $\pi_{i}$ as input, and outputs the action $a_{i}$. (Note that the Actor Network outputs real numbers, and we round them to the nearest integers to get the action $a_{i}$.) After an iteration, the $N$ local states $(\pi_{1}, \pi_{2}, \cdots, \pi_{N})$, the $N$ actions $(a_{1}, a_{2}, \cdots, a_{N})$ and the overall reward $R$ of the iteration are stored in a buffer. The buffer has a fixed size. When new data come in, if the buffer is full, the oldest data will be removed. Therefore, the buffer always stores the most recent results.

After each iteration, a number of samples will be randomly chosen from the buffer to train the four networks. Each sample has the form of $(\pi_{i}, a_{i}, \pi_{i+1}, R)$. The four networks update their weights as follows, using the idea of the DDPG algorithm~\cite{lillicrap2015continuous}:
\begin{itemize}
    \item Step 1: \emph{train the Critic Network}. As shown in Figure~\ref{fig:actor_critic}, the Critic Network takes $\pi_{i}$ and $a_{i}$ as input, and outputs a value $Q(\pi_{i}, a_{i})$. We also concatenate the Target Actor Network and the Target Critic Network (as shown in Figure~\ref{fig:actor_critic}), and use $\pi_{i+1}$ as input to generate the output $Q^{target}(\pi_{i+1}, a_{i+1}^{target})$. The loss function of the Critic Network is then set as 
    \begin{equation}
    \mathcal{L}_{critic} = (Q(\pi_{i}, a_{i}) - \gamma Q^{target}(\pi_{i+1}, a_{i+1}^{target}) - (R - \mathcal{B}))^{2}
    \end{equation}
    where the baseline $\mathcal{B}$ is defined as an exponential moving average of all previous rewards in order to reduce the variance of gradient estimation. A small number of samples are used as a mini-batch, and their total loss is used to update the weights of the Critic Network via backpropagation.
    \item Step 2: \emph{train the Actor Network}. We concatenate the Actor Network and the Critic Network (as shown in Figure~\ref{fig:actor_critic}), and use $\pi_{i}$ to generate the output $Q(\pi_{i}, a_{i})$. The loss function is then set as 
    \begin{equation}
        \mathcal{L}_{actor} = - Q(\pi_{i}, a_{i})
    \end{equation}
    Then the total loss of a mini-batch of such samples is used to update the weights of the Actor Network via backpropagation (with the weights of the Critic Network frozen).
    \item Step 3: \emph{train the Target Actor Network}. Let $\delta$ be a small number, such as $\delta = 0.01$. Let $w_{actor}^{target}$ be a weight of the current Target Actor Network, and let $w_{actor}$ be the corresponding weight of the updated Actor Network. We update $w_{actor}^{target}$ as:
    \begin{equation}
        w_{actor}^{target} \leftarrow w_{actor}^{target} + \delta(w_{actor} - w_{actor}^{target})
    \end{equation}
    We update all weights of the Target Actor Network in the same way.
    \item Step 4: \emph{train the Target Critic Network}. We update its weights in the same way as we did with the Target Actor Network, except that here we consider the Target Critic Network and the Critic Network.
\end{itemize}

In summary, the Critic Network learns to predict the future rewards given the current state and the action to be taken. The Actor Network learns to take the best action based on the future rewards predicted by the Critic Network. The Target Critic Network (respectively, the Target Actor Network) follows the learning of the Critic Network (respectively, the Actor Network), except that it updates its weights at a slower pace, which is a conservative method that helps the DRL algorithm converge. The DRL algorithm ends when the four networks' performance converges or when a preset number of training steps is reached. 

\section{Experimental Evaluation and Analysis}
\label{sec:exp}
In this section, we present experimental evaluation of the Selected Protection scheme. We focus on two important deep neural networks in computer vision: ResNet-18~\cite{DBLP:journals/corr/HeZRS15} and VGG16~\cite{simonyan2014very}. We consider two well-known datasets: the CIFAR-10 dataset~\cite{cifar} and the MNIST dataset~\cite{726791}. We use two data representation schemes for the weights: the IEEE-754 floating-point representation, and the fixed-point representation. We explore two types of error correcting codes: an ideal ECC that reaches the Shannon capacity, and a practical finite-length BCH code. And we study the performance of two methods for the SP scheme: the \BitMask{} method and the \TopBits{} method.

The experimental results show that the Selective Protection scheme based on deep reinforcement learning can substantially outperform the natural baseline scheme, where all layers protect the same number of bits. The experimental results also reveal a very interesting fact: the Most Significant Bits (MSBs) in a data representation do not always affect the performance of a neural network in the most significant ways. Consequently, the \BitMask{} method can sometimes protect some less significant bits (instead of MSBs) and outperform the \TopBits{} method. We present a detailed analysis of this surprising finding.

In the following, we introduce the setup of experiments, and present the redundancy-performance tradeoff of the SP scheme. We then show how the \BitMask{} method and the \TopBits{} method select bits for protection, and analyse why sometimes LSBs are more important for the performance of neural networks than MSBs in noisy environments.

\subsection{Setup of Experiments}
We test the performance of the SP scheme on two important neural network models: ResNet-18 and VGG16. Both models are commonly used for classifying images, and have various applications in computer vision. The architectures of the two models are illustrated in Figure~\ref{fig:network}. The ResNet-18 network has 26 edge layers and 11.69 million weights. The VGG16 network has 16 edge layers and 138 million weights. Such sizes are typical for deep neural networks.

We perform image classification tasks on two important datasets: the CIFAR-10 dataset and the MNIST dataset. The CIFAR-10 dataset consists of 60, 000 colored images of size $32 \times 32$ each, which belong to 10 different classes. The MNIST dataset consists of 70, 000 gray-scaled images of size $28 \times 28$ each, which represent the 10 classes of hand-written digits from 0 to 9. Both datasets are widely used for testing the performance of image classification.

We study the SP scheme for two data representation methods: the IEEE-754 floating-point representation and the fixed-point representation. The IEEE-754 representation is an international standard widely used in most hardware systems. The fixed-point representation is a natural alternative way to quantize weights with easily controllable ranges and quantization precision. In our experiments, we let the IEEE-754 representation use 32 bits for each weight, and let the fixed-point representation use 8 bits for each weight.

We explore two types of ECCs for protecting the important bits selected by the SP scheme. The first one is an ideal ECC that reaches the Shannon capacity. When the weights suffer from errors of a binary symmetric channel with BER $p$, we let the ideal ECC have a code rate of $1-H(p)$, matching the channel's capacity. We use the code to protect all the selected important bits, and assume that decoding always succeeds. The second type of codes are practical finite-length BCH codes. When the IEEE-754 floating-point representation is used, we let the code be a (8191, 6722) BCH code, which can correct 115 errors. When the fixed-point representation is used, we let the code be a (8191, 6787) BCH code, which can correct 110 errors. When $p = 0.01$ (a practical BER for storage systems), both codes can decode with sufficiently small failure probabilities, thus causing minimal degradation for the neural network's performance.

We study the performance of two methods for the SP scheme: the \BitMask{} method and the \TopBits{} method. The \BitMask{} method offers greater freedom in selecting which bits to protect, while the \TopBits{} method offers higher efficiency for learning due to its more restricted solution space. For both methods, the deep reinforcement learning algorithm converges efficiently. Given a solution of the SP scheme, we generate random errors 100 times for all the weights, and evaluate the neural network's average performance (i.e. classification accuracy). The performance was found to be stable over different experiments.

\subsection{Redundancy-Performance Tradeoff}

\begin{figure}[!t]
    \centering
    \begin{subfigure}[b]{0.45\textwidth}
        \includegraphics[width=\textwidth]{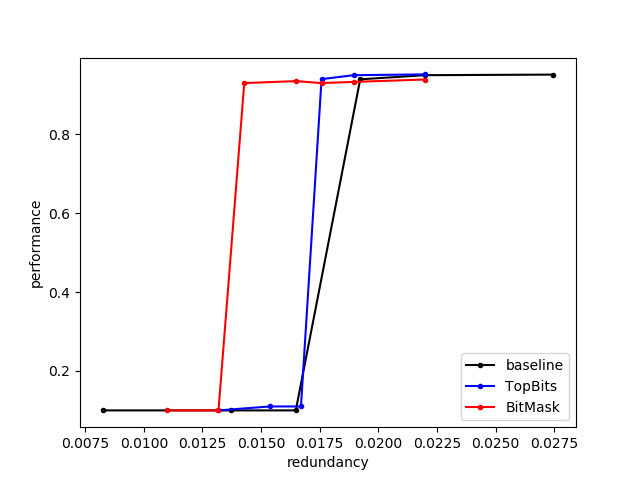}
        \caption{}
    \end{subfigure}
    \hfill
    \begin{subfigure}[b]{0.45\textwidth}
        \includegraphics[width=\textwidth]{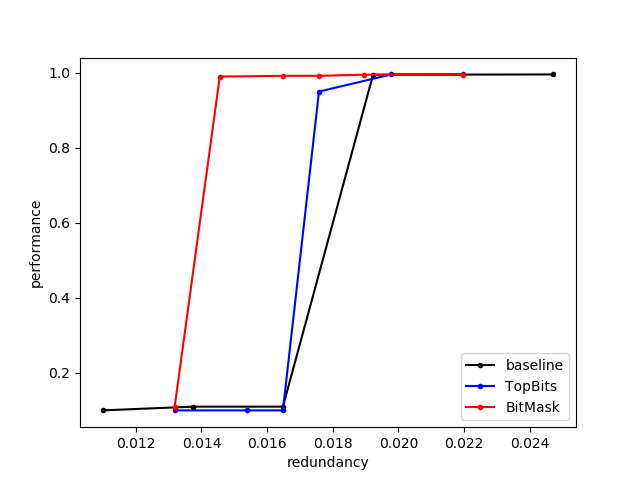}
        \caption{}
    \end{subfigure}
    
    \begin{subfigure}[b]{0.45\textwidth}
        \includegraphics[width=\textwidth]{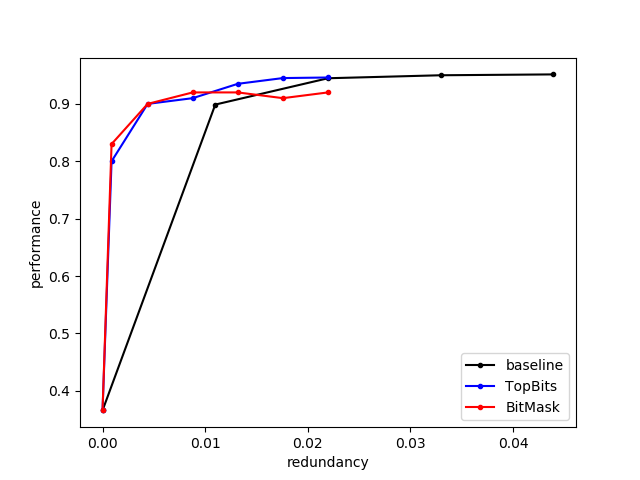}
        \caption{}
    \end{subfigure}
    \hfill
    \begin{subfigure}[b]{0.45\textwidth}
        \includegraphics[width=\textwidth]{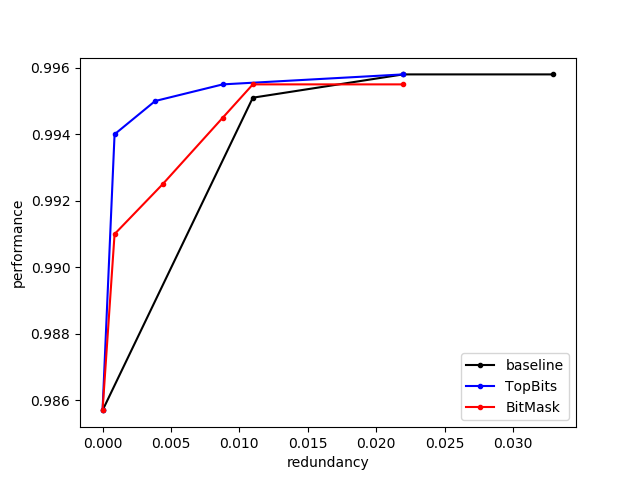}
        \caption{}
    \end{subfigure}
    \caption{The redundancy-performance tradeoff for the SP scheme when ideal ECC is used. Here ``baseline'', ``\TopBits{}'' and ``\BitMask{}'' denote the baseline algorithm (where all layers protect the same number of bits), the \TopBits{} method and the \BitMask{} method, respectively. (a) The neural network is ResNet-18, the dataset is CIFAR-10, and the data representation scheme is IEEE-754. (b) The neural network is VGG16, the dataset is MNIST, and the data representation scheme is IEEE-754. (c) The neural network is ResNet-18, the dataset is CIFAR-10, and the data representation scheme is fixed-point. (d) The neural network is VGG16, the dataset is MNIST, and the data representation scheme is fixed-point.}
    \label{fig:ideal_result}
\end{figure}

\begin{figure}[!t]
    \centering
    \begin{subfigure}[b]{0.45\textwidth}
        \includegraphics[width=\textwidth]{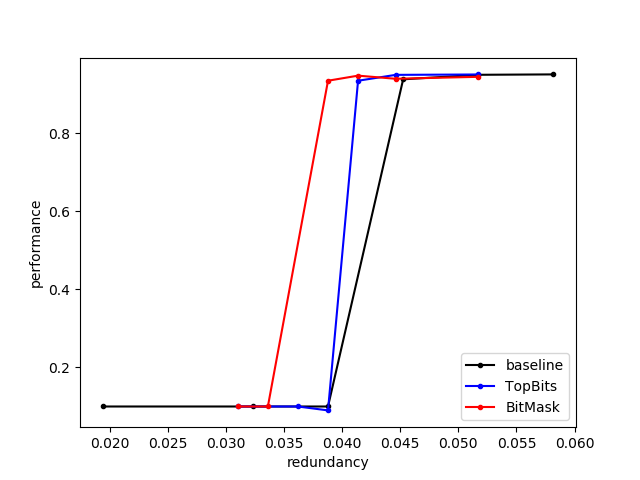}
        \caption{}
    \end{subfigure}
    \hfill
    \begin{subfigure}[b]{0.45\textwidth}
        \includegraphics[width=\textwidth]{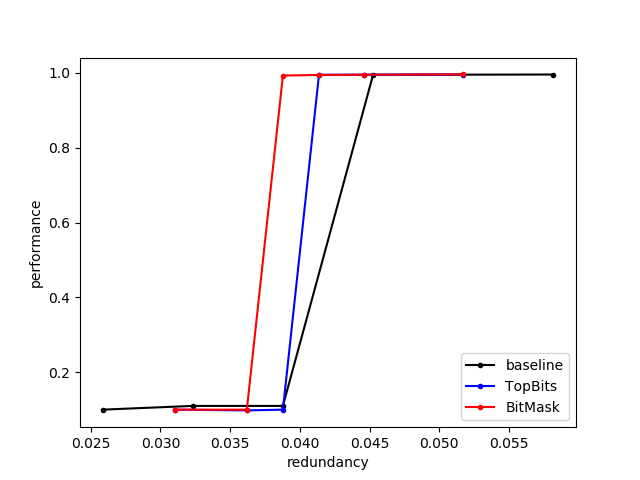}
        \caption{}
    \end{subfigure}
    
    \begin{subfigure}[b]{0.45\textwidth}
        \includegraphics[width=\textwidth]{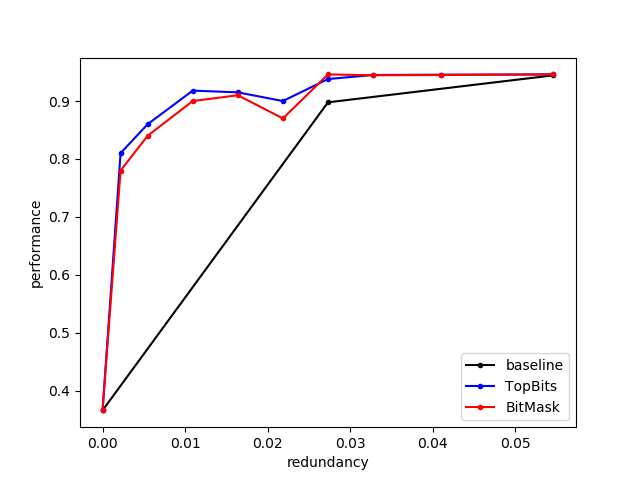}
        \caption{}
    \end{subfigure}
    \hfill
    \begin{subfigure}[b]{0.45\textwidth}
        \includegraphics[width=\textwidth]{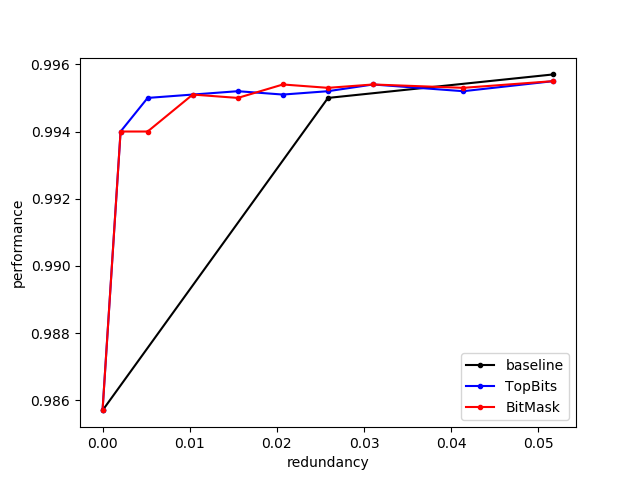}
        \caption{}
    \end{subfigure}
    \caption{The redundancy-performance tradeoff for the SP scheme when BCH codes are used. (a) The neural network is ResNet-18, the dataset is CIFAR-10, and the data representation scheme is IEEE-754. (b) The neural network is VGG16, the dataset is MNIST, and the data representation scheme is IEEE-754. (c) The neural network is ResNet-18, the dataset is CIFAR-10, and the data representation scheme is fixed-point. (d) The neural network is VGG16, the dataset is MNIST, and the data representation scheme is fixed-point.}
    \label{fig:bch_result}
\end{figure}

The experimental results for the \emph{redundancy-performance tradeoff} are shown in Figure~\ref{fig:ideal_result} and Figure~\ref{fig:bch_result}. They are for two different types of ECCs, respectively: Figure~\ref{fig:ideal_result} is for the ideal ECC, while Figure~\ref{fig:bch_result} is for the finite-length BCH codes. In all experiments, we let BER be $p=0.01$. The redundancy $r = \frac{k_{pro} (n-k)}{k_{total} k}$ can be adjusted by setting different target redundancy in the deep reinforcement learning algorithm. The performance is measured as the average classification accuracy of the neural network, whose noisy weights are partially protected by the ECC.

The figures show that when the redundancy $r$ is relatively large, the neural network retains its high performance (because the bits most important for its performance are protected by ECCs). However, once the redundancy drops below a certain threshold, the performance drops sharply. It can be seen clearly that, overall, both the \BitMask{} method and the \TopBits{} method \emph{significantly} outperform the baseline method, where all layers protect the same number of bits. (In the baseline method, we always protect the first few bits in the weights because they are more significant.)

It can also be seen that when the IEEE-754 representation is used, the \BitMask{} method outperforms the \TopBits{} method substantially overall. When the fixed-point representation is used, the performance of two methods becomes more comparable, with the \TopBits{} method sometimes outperforming the \BitMask{} method. It is a very interesting observation because the \TopBits{} method always chooses the first few bits of each weight, which are usually considered more significant than the remaining bits. Furthermore, this restriction also reduces the dimensions of the solution space substantially, which helps improve the efficiency of learning. It implies that the \BitMask{} method can find less significant bits that are more important than MSBs for a neural network's overall performance. In the following, we analyse this surprising result by studying how the two methods select bits, and how the bits affect the neural network's performance.

\subsection{Bits Protected by Selective Protection Scheme}

\begin{figure}[!t]
    \centering
    \begin{subfigure}[b]{0.7\textwidth}
        \includegraphics[width=\textwidth]{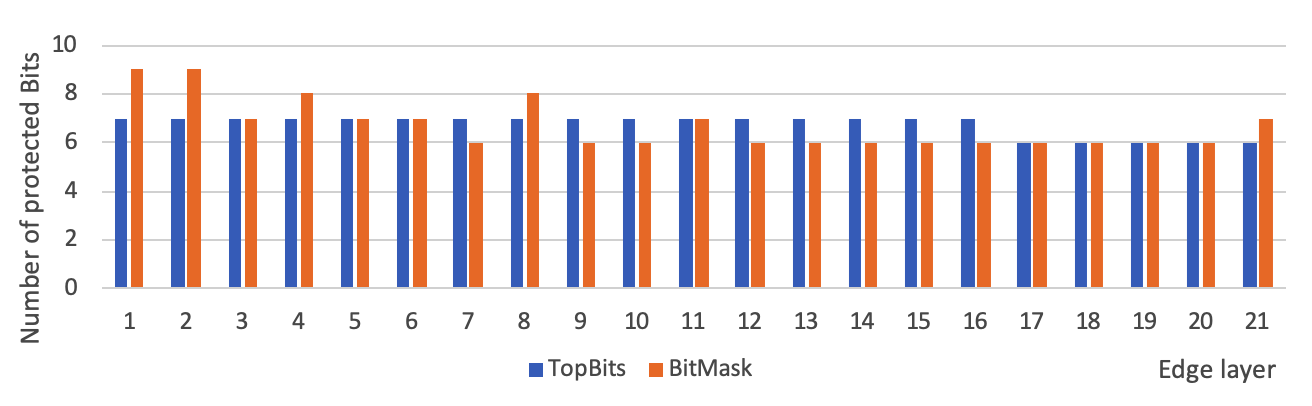}
        \caption{IEEE-754 floating-point representation}
    \end{subfigure}
    \hfill
    \begin{subfigure}[b]{0.77\textwidth}
        \includegraphics[width=\textwidth]{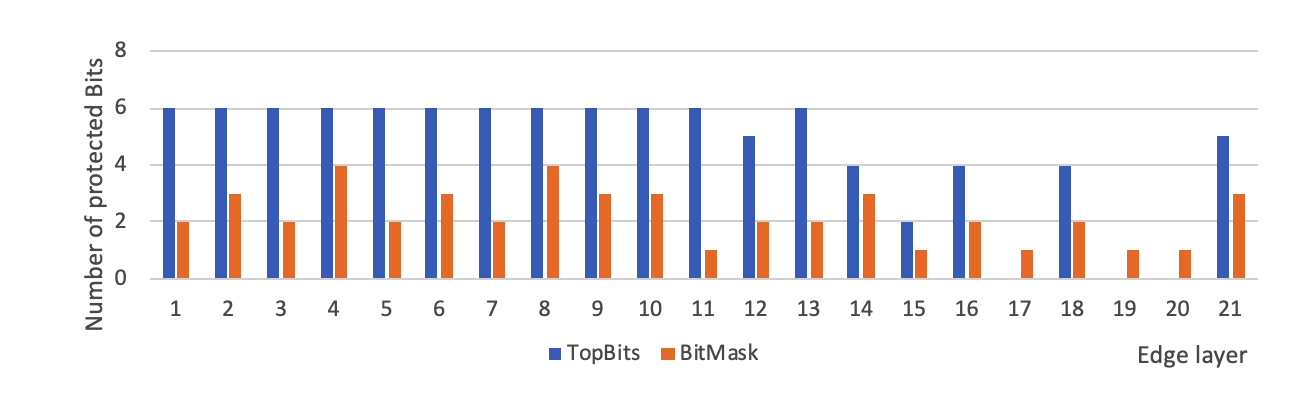}
        \caption{Fixed-point representation}
    \end{subfigure}
    \caption{The number of selected bits for ECC protection in each edge layer. Here the neural network is ResNet-18, the dataset is CIFAR-10, and the ECC is the ideal ECC. Sub-figure (a) is for the IEEE-754 floating-point representation, and sub-figure (b) is for the fixed-point representation. The orange bars are for the \BitMask{} method (with the redundancy $r = 0.189$ for (a) and $0.193$ for (b)) , and the blue bars are for the \TopBits{} method (with the redundancy $r = 0.2$ for (a) and $0.2$ for (b)). (Note that both methods used the same target redundancy $\bar{r}$ in their DRL algorithm. But since the DRL algorithm only makes the final redundancy be close to the target redundancy, their final redundancies are not identical.)}
    \label{fig:bit_prob_1}
\end{figure}

We now study how the \emph{BitMask} method and the \emph{TopBits} method select bits. For the \emph{number} of bits selected by the two methods, its distribution over the layers is as illustrated in Figure~\ref{fig:bit_prob_1}. It can be seen that when the data representation is IEEE-754, both methods have a relatively even distribution over the layers. And when the data representation is the fixed-point representation, the distribution for both methods becomes less even. Overall, the two methods behave similarly in this aspect. 

\begin{figure}[!t]
    \centering
    \includegraphics[width=0.6\textwidth]{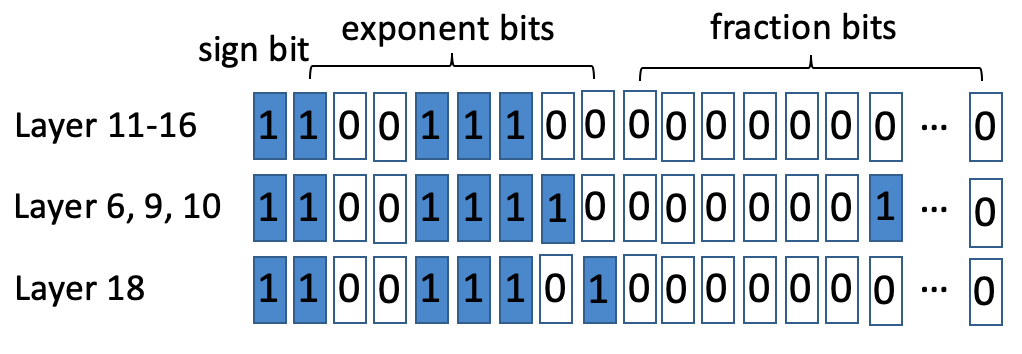}
    \caption{Typical examples of the bit-mask vector in some edge layers, with the IEEE-754 floating-point representation and the \BitMask{} method. Here the neural network is ResNet-18, the dataset is CIFAR-10 and the ECC is the ideal ECC. The positions of the selected bits for ECC protection correspond to the 1's in the bit-mask vector (of the blue color). Notice that among the exponent bits, some less significant bits are selected instead of more significant bits.}
    \label{fig:resnet_float_pattern}
\end{figure}

The major difference between the \BitMask{} method and the \TopBits{} method is in \emph{which} bits they select. Since their redundancy-performance tradeoff differs most significantly when the IEEE-754 representation is used, we focus on the IEEE-754 representation from now on. For the \TopBits{} method, it always selects the first few bits in each layer. For the \BitMask{} method, however, it selects bits quite differently. Some typical examples are shown in Figure~\ref{fig:resnet_float_pattern}. It shows that instead of selecting some more significant bits (such as the third and the fourth bits), the \BitMask{} method selects some less significant bits (such as the fifth, sixth and seventh bits in the $11$th layer, the $12$th layer, $\cdots$, the $16$th layer). The result is intriguing because the more significant bits affect the value of a weight more substantially, and are usually expected to affect the performance of the neural network more as well. We present the analysis for the result in the next subsection.

\subsection{Analysis of BitMask Method and TopBits Method}

When the IEEE-754 data representation is used, consider a bit among the exponent bits. (The exponent bits are where the \BitMask{} method's selection and the \TopBits{} method's selection differ the most. For fraction bits, most of them are not selected by either method.) Recall that for a weight, when its bits are $(b_{0}, b_{1}, \cdots, b_{31})$, the corresponding weight is $w = (-1)^{(b_{0})_{2}} \times 2^{(b_{1}b_{2}\cdots b_{8})_{2}-127} \times (1.b_{9}b_{10} \cdots b_{31})_{2}$. Let $1 \leq i \leq 8$, and consider the exponent bit $b_{i}$. There are two important factors that determine how an error in $b_{i}$ affects the neural network's performance:
\begin{enumerate}
    \item Factor one: The 0-to-1 error and the 1-to-0 error have an asymmetric impact on the neural network's performance.
    \item Factor two: The bit $b_{i}$ can have a highly imbalanced probability distribution, which also affects the performance.
\end{enumerate}

We analyze the two factors in the following. For the first factor, consider a 0-to-1 error that changes bit $b_{i}$ from 0 to 1. In this case, the weight changes from $w$ to $w_{0-to-1} = 2^{2^{8-i}} \times w$. With a 1-to-0 error that changes the bit $b_{i}$ from 1 to 0, the weight will change from $w$ to $w_{1-to-0} = 2^{-2^{8-i}} \times w$. Since each neuron takes a linear combination of its incoming values before passing it to an activation function, the absolute value of the weight plays an important role in the function of the neuron. It is easy to see that the $0-to-1$ error changes the absolute value of the weight much more significantly than the $1-to-0$ error. So the $0-to-1$ errors are expected to affect the neural network's performance more significantly as well.

\begin{figure}[!t]
    \begin{subfigure}[b]{0.45\textwidth}
    \centering
    \includegraphics[width=\textwidth]{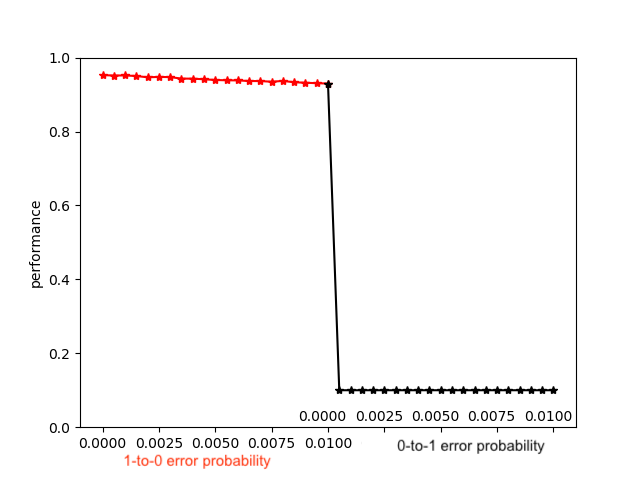}
    \caption{}
    \end{subfigure}
    \hfill
    \begin{subfigure}[b]{0.45\textwidth}
    \centering
    \includegraphics[width=\textwidth]{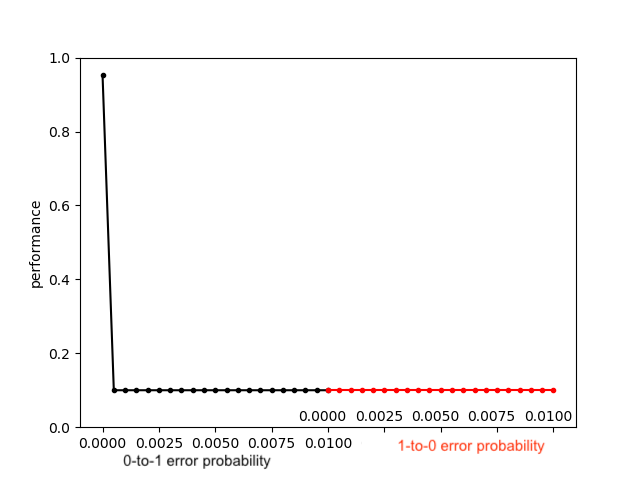}
    \caption{}
    \end{subfigure}
    \hfill
    \begin{subfigure}[b]{0.45\textwidth}
    \centering
    \includegraphics[width=\textwidth]{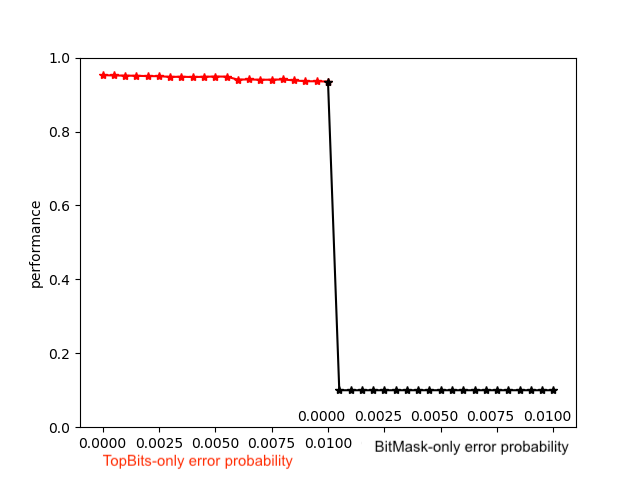}
    \caption{}
    \end{subfigure}
    \hfill
    \begin{subfigure}[b]{0.45\textwidth}
    \centering
    \includegraphics[width=\textwidth]{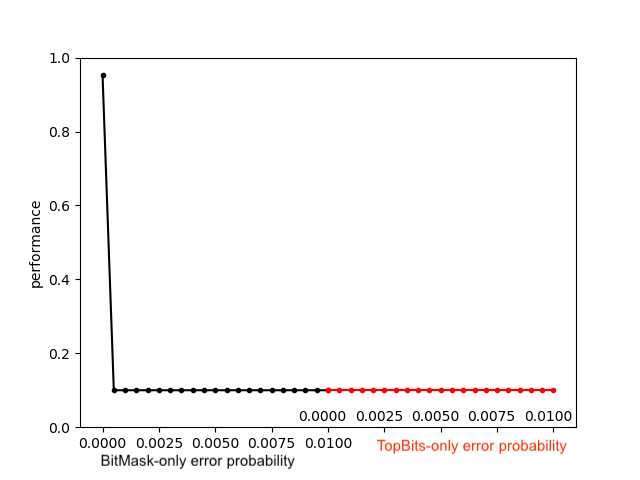}
    \caption{}
    \end{subfigure}
    \caption{How the performance of a neural network changes when errors are added to its bits in two phases. (No bits here are protected by ECC.) Here the neural network is ResNet-18, the dataset is CIFAR-10, and the data representation is IEEE-754. (a) In phase 1, we only add 1-to-0 errors to bits that are originally 1s; then in phase 2, we continue to add 0-to-1 errors to bits that are originally 0s. In both phases, we gradually increase the error probability from 0 to 1\% (and the same holds for the other three sub-figures). (b) In phase 1, we only add 0-to-1 errors to bits that are originally 0s; then in phase 2, we continue to add 1-to-0 errors to bits that are originally 1s. (c) Let $S_{TopBits}$ be the set of bits selected by the \TopBits{} method, and let $S_{BitMask}$ be the set of bits selected by the \BitMask{} method. In phase 1, we only add errors to the bits that are in the set $S_{TopBits} - S_{BitMask}$; then in phase 2, we continue to add errors to the bits that are in the set $S_{BitMask} - S_{TopBits}$. (d) In phase 1, we only add errors to the bits that are in the set $S_{BitMask} - S_{TopBits}$; then in phase 2, we continue to add errors to the bits that are in the set $S_{TopBits} - S_{BitMask}$.}
    \label{fig:error}
\end{figure}

We experimentally verify the above observation in Figure~\ref{fig:error} (a) and (b). They show that when 0-to-1 errors are added, the performance of the neural network drops very sharply. When 1-to-0 errors are added, however, the performance of the neural network does not change much. The results verify that 0-to-1 errors have a more significant impact on the neural network's performance. So to achieve an optimal redundancy-performance tradeoff, there is a strong motivation to protect bits that are more likely to be 0s.

\begin{figure}[!t]
    \centering
    \includegraphics[width=\textwidth]{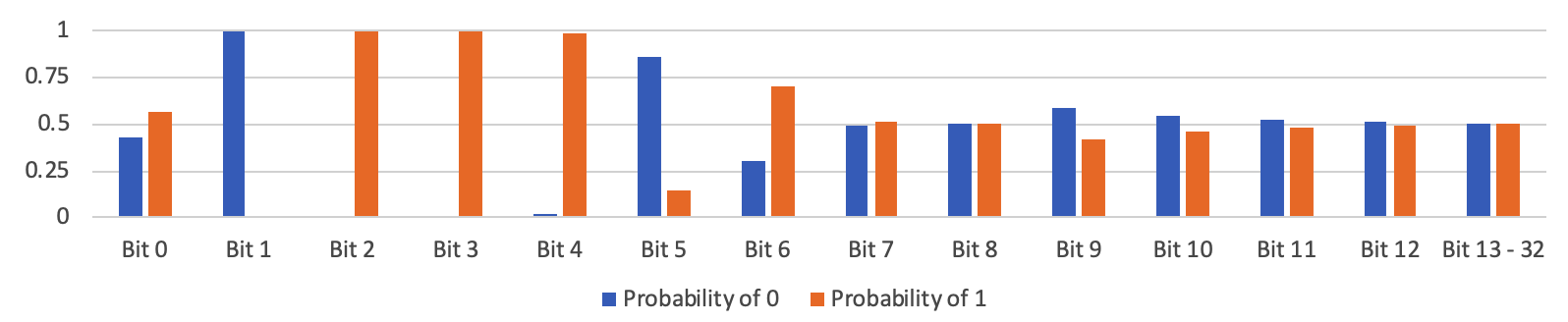}
    \caption{The probability distribution of the bits in each bit position. Here the neural network is ResNet-18, the dataset is CIFAR-10, and the data representation is the IEEE-754 standard. The $x$-axis shows the 32 bit positions for weights. The $y$-axis shows the probability for a bit in each position to be 0 or 1. (Each blue bar is the probability for the bit to be 0, and each orange bar is the probability for the bit to be 1. Here the weights are noiseless.) It can be seen that for some exponent bits (especially from bit 1 to bit 6), the probablity distribution can be quite uneven. (The orange bar for bit 1 and the blue bar for bit 2 have height 0, and therefore cannot be seen.)}
    \label{fig:bit_prob}
\end{figure}

Let us now study the probability distribution of the bits in each bit position. The results are as illustrated in Figure~\ref{fig:bit_prob}. It can be seen that for many exponent bits (including bit 1 to bit 6), the probability distribution can be quite uneven. In fact, due to the weight distribution in the neural network, bit 2 and bit 3 here are nearly always 1s, and that explains why they were not selected by the \BitMask{} method (as shown in Figure~\ref{fig:resnet_float_pattern}). Overall, whether a bit should be selected depends on the balance between both factors: the level of asymmetry in the impact on performance by the 0-to-1 errors and the 1-to-0 errors, and the probability for the bit to be 0 or 1. The greater the level of asymmetry is, and the more probable the bit is 0, the more likely the bit will be selected.

We study the bits that are selected differently by the \BitMask{} method and the \TopBits{} method, and explore their impact on the neural network's performance.  The experimental results are shown in Figure~\ref{fig:error} (c) and (d). Let $S_{TopBits}$ be the set of bits selected by the \TopBits{} method, and let $S_{BitMask}$ be the set of bits selected by the \BitMask{} method. (Here we let the \TopBits{} method select the same number of bits as the \BitMask{} method in each layer for fair comparison.) It can be seen that when errors are added to the bits in $S_{BitMask} - S_{TopBits}$, the performance of the neural network drops very sharply. When errors are added to the bits in $S_{TopBits} - S_{BitMask}$, however, the performance does not change much. The results verify that the \BitMask{} method indeed chooses bits that are more important for the redundancy-performance tradeoff.


\section{Conclusions}
\label{sec:con}
In this work, we use deep learning to selectively protect the weights in neural networks from errors, in order to achieve an optimized redundancy-performance tradeoff. The error-correction scheme is function-oriented: it aims at optimizing the neural network's overall performance, instead of the uncorrectable bit error rates among all the bits after decoding. It studies two important methods for the Selective Protection scheme: the \BitMask{} method and the \TopBits{} method. Both methods outperform the baseline scheme significantly. And interestingly, it was discovered that sometimes, protecting less significant bits (LSBs) is more important to the neural network's performance than protecting some more significant bits (MSBs).

The proposed error-correction paradigm can be extended in various ways. One interesting extension is to study how errors in different modules in a neural network (including filters, channels, attention modules, \emph{etc}.) affects the neural network's performance, and design error-correction schemes accordingly. They remain as our future research.

\bibliographystyle{unsrt}
\bibliography{fec_rnn_ref}
\end{document}